\documentclass[conference]{IEEEtran}
\IEEEoverridecommandlockouts

\usepackage{cite}
\usepackage{amsmath,amssymb,amsfonts}
\usepackage{algorithmic}
\usepackage{graphicx}
\usepackage{textcomp}
\usepackage{booktabs}
\usepackage[dvipsnames]{xcolor}
\usepackage{hyperref}
\hypersetup{
colorlinks,
linkcolor=NavyBlue,
citecolor=NavyBlue,
urlcolor=SeaGreen
}

\usepackage[a4paper, total={184mm,239mm}]{geometry}
\def\BibTeX{{\rm B\kern-.05em{\sc i\kern-.025em b}\kern-.08em
    T\kern-.1667em\lower.7ex\hbox{E}\kern-.125emX}}

\usepackage{algorithm}
\usepackage{multirow,multicol} 
\usepackage{tabulary} 
\usepackage{tabularx} 
\usepackage{makecell} 
\usepackage{enumerate,enumitem}
\usepackage{graphicx}
\usepackage{float} 
\usepackage{colortbl}

\begin{document}

\title{Generalized Ping-Pong: Off-Chip Memory Bandwidth Centric Pipelining Strategy for Processing-In-Memory Accelerators}

\author{\IEEEauthorblockN{Ruibao Wang$^{1,2}$, and Bonan Yan$^{1*}$}
\IEEEauthorblockA{$^1$Institute for Artificial Intelligence, Peking University, Beijing, China\\$^2$College of Electronic Science \& Engineering, Jilin University
\\Email: wangrb1921@mails.jlu.edu.cn, bonanyan@pku.edu.cn}
}

\maketitle

\begin{abstract}
Processing-in-memory (PIM) is a promising choice for accelerating deep neural networks (DNNs) featuring high efficiency and low power. However, the rapid upscaling of neural network model sizes poses a crucial challenge for the limited on-chip PIM capacity. When the PIM presumption of ``pre-loading DNN weights/parameters only once before repetitive computing'' is no longer practical, concurrent writing and computing techniques become necessary for PIM. Conventional methods of naive ping-pong or in~situ concurrent write/compute scheduling for PIM cause low utilization of off-chip memory bandwidth, subsequently offsetting the efficiency gain brought by PIM technology. To address this challenge, we propose an off-chip memory bandwidth centric pipelining strategy, named ``generalized ping-pong'', to maximize the utilization and performance of PIM accelerators toward large DNN models. The core idea of the proposed generalized ping-pong strategy is to evenly distribute the active time and fully utilize the off-chip memory bandwidth. Based on a programmable and scalable SRAM PIM architecture, we quantitatively analyze and compare the generalized ping-pong with the conventional scheduling strategies of naive ping-pong and in~situ write/compute for PIM. Experiments show that the generalized ping-pong strategy achieves acceleration of over 1.67$\times$ when fully utilizing the off-chip memory bandwidth. When further limiting the off-chip memory bandwidth ranging in 8$\sim$256 bytes per clock cycle,
the proposed generalized ping-pong strategy accelerates 1.22$\sim$7.71$\times$ versus naive ping-pong. The developed PIM accelerator design with the generalized ping-poing strategy is open-sourced at \url{https://github.com/rw999creator/gpp-pim}.
\end{abstract}


\section{Introduction}
Processing-In-Memory (PIM) is an innovative approach that holds the potential to significantly accelerate deep learning operations by enabling computations within memory arrays rather than transferring data back and forth between processing units and storage mediums~\cite{zhou2023overlapim}.
The fundamental concept of PIM revolves around integrating computing circuits within or near memory arrays to process data directly on stored information, especially vector-matrix multiplication~\cite{jiang2024gnndrive,radway2021illusion}. However, deep neural network (DNN) models, following the deep learning scaling law, are scaling up at an exponential speed~\cite{zhang2018caffeine,dos202412nm,kelefouras2023design,hager202411,alverti2022daxvm}, inflicting unprecedented challenges for the limited on-chip PIM capacity in that most of the conventional PIM architectures hold the presumption that loading weights (parameters) of deep learning models only once before repetitive computation based on a weight-stationary parallelism scheme~\cite{perri2024digital}. 

\begin{figure}[t]
\includegraphics[width=1\columnwidth]{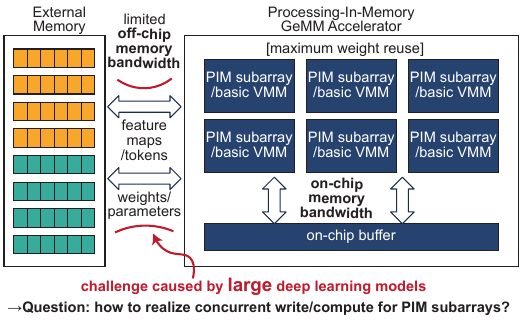}
\vspace{-1.5em}
\caption{PIM accelerators face emerging challenge in high-dimensional GeMM computation: realization of concurrent write/compute.}
\label{fig:1}
\vspace{-1.5em}
\end{figure}

In contrast, the trending largest ever deep learning models (e.g. Transformer-based large language models~\cite{yang2020retransformer,kashikar2023lossless} and large multimodal models~\cite{zhuang2024ssr}) have required reloading DNN weights in PIM architectures into a necessary feature~\cite{razavi2022exploiting}.
The weights are sliced and programmed to the PIM subarray (i.e. macros) in batches amid PIM computation for general matrix multiplication (GeMM), i.e. concurrent write/compute (Fig.~\ref{fig:1}).

There are two existing practical strategies to schedule the weight writing and PIM computation~\cite{yue202014,aranda2020analysis}.
(a) \textbf{In~situ write/compute strategy} stalls the computation of PIM macro for rewriting new weights. This strategy aims to synchronize all PIM macros for rewriting and computing, which degrades the utilization rate and subsequently offsets the performance gain brought by PIM. 
(b) \textbf{Naive ping-pong strategy} facilitates concurrent write/compute pairing of two PIM macros, where one is computing and the other is updating weights. Although the naive ping-pong strategy hides the weight rewriting delay, it hardly balances the two operations, causing potential pipeline bubbles and low utilization. In other words,the compute time and weight updating time are different in most cases, resulting some PIM macros keep waiting for the other ones. 
We observe that the existing two common strategies have certain limitations and are primarily focused on the PIM chip itself, hardly taking into account the impact of off-chip memory bandwidth on the overall performance.

To address this challenge, we propose an off-chip memory bandwidth centric pipelining strategy, named ``\textbf{generalized ping-pong}'', to maximize the utilization and performance of PIM accelerators toward large DNN models. The core idea of the proposed generalized ping-pong strategy is to evenly distribute the active time and fully utilize the off-chip memory bandwidth to achive high utilization of PIM arrays and off-chip memory bandwidth. Moreover, we tailor a scalable PIM architecture equipped with an assembler and customized instruction set, thereby analyzing metrics such as execution time, peak bandwidth requirements, and macro utilization across different strategies. The developed PIM accelerator design with the generalized ping-poing strategy is open-sourced at \url{https://github.com/rw999creator/gpp-pim}.

Experiments show that the generalized ping-pong strategy achieves an acceleration of over 1.67$\times$ when fully utilizing the off-chip memory bandwidth. When further limiting the off-chip memory bandwidth ranging from 8$\sim$256 bytes per clock cycle, the proposed generalized ping-pong strategy accelerates 1.22$\sim$7.71$\times$ versus the existing naive ping-pong strategy.

\section{Preliminaries}
\subsection{SRAM-Based PIM Designs}
\begin{figure}[t]
\includegraphics[width=1\columnwidth]{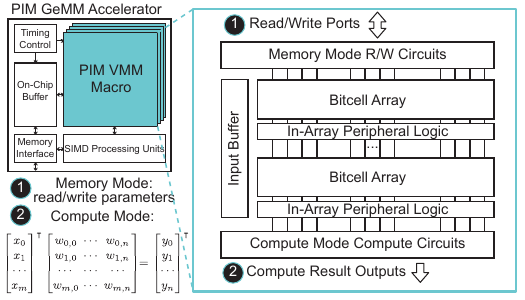}
\vspace{-1.5em}
\caption{Typical SRAM-based PIM GeMM accelerator circuit block diagram and its two work modes.}
\label{fig:pim-macro}
\vspace{-1em}
\end{figure}

PIM GeMM accelerator consists of multiple PIM vector-matrix multiplication (VMM) macros (subarrays) to perform complete GeMM operations (Fig.~\ref{fig:pim-macro})~\cite{tang2024pimlc,kim20231,yan-sram,wang202434}. Each PIM macro works in two primary operational modes: memory model and compute mode~\cite{fu2023probabilistic}. The memory mode serves a crucial role in loading weights/parameters into the PIM macro for maximum reuse in the compute mode. The compute mode is dedicated to performing in-memory vector-matrix multiplication (VMM) computations that leverage the physical locality of data within SRAM bitcells. Static Random Access Memory (SRAM)-based PIM offers both fast computing speed in the compute mode and low read/write latency in the memory mode.  Also, SRAM is more appropriate for repetitively reloading with over 10$^{15}$ bitcell endurance. However, the density of SRAM-PIM leads to limited on-chip capacity (e.g. 16kb$\sim$4.5Mb/macro). Toward the upscaled deep learning models, concurrent write/compute strategies is in urgent need for SRAM-based PIM.

\subsection{Existing Concurrent Write/Compute Strategies}

\begin{figure*}[t]
\includegraphics[width=1\linewidth]{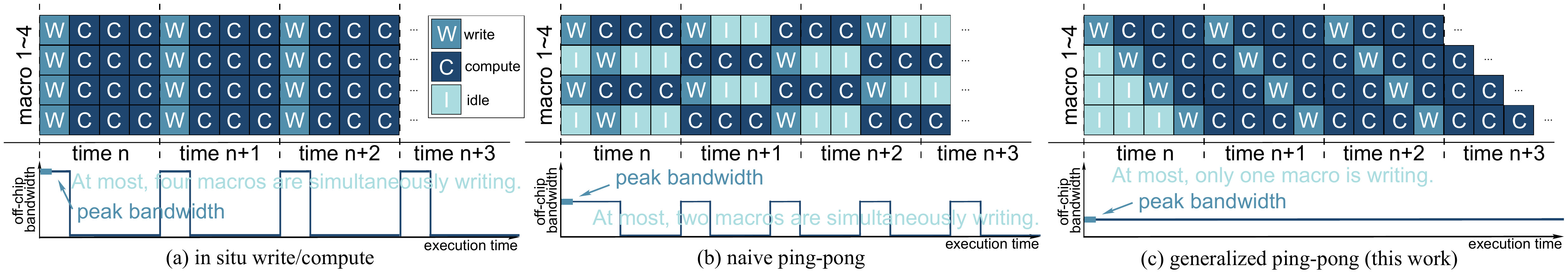}
\vspace{-1.5em}
\caption{(a) In situ write/compute strategy. (b) Naive ping-pong strategy. (c) Generalized ping-pong strategy (this work).}
\label{fig:schedulescheme}
\vspace{-1em}
\end{figure*}

Fig.~\ref{fig:schedulescheme} illustrates the comparison between different concurrent write/compute strategies using an exemplary PIM accelerator comprising 4 PIM macros.
Fig.~\ref{fig:schedulescheme}(a) illustrates the in~situ write/compute strategy. It synchronizes all PIM macros for writing or computing.
Only writing occupies the off-chip memory bandwidth, reflecting an intermittent characteristic.
Fig.~\ref{fig:schedulescheme}(b) depicts the naive ping-pong strategy~\cite{liu2018parallelizing}.
With $>$2 PIM macros, the naive ping-pong strategy divides all macros into two groups, say, \textit{bank1} and \textit{bank2}. While \textit{bank1} performs computations for the n$^{\mathrm{th}}$ GeMM operation, \textit{bank2} loads the weights for the (n+1)$^{\mathrm{th}}$ opeartion; once the computations for the n$^{\mathrm{th}}$ operation are completed, \textit{bank1} loads the weights for the (n+1)$^{\mathrm{th}}$, and \textit{bank2} executes computations for the (n+1)$^{\mathrm{th}}$ operation. 
This partitioning of computation and rewriting areas is achieved through two methods: inter-macro ping-pong~\cite{yue202014}, which partitions between macros, and intra-macro ping-pong, which partitions within a macro~\cite{adve2024agile,song2019hypar,grimm2022neural,hao2019fpga,chen2019cloud}.
It alleviates the utilization for off-chip memory bandwidth, but idle time still exists depending on the comparison between intrinsic PIM macro computing throughput and data capacity~\cite{jiang2023acgraph,chen2023hardware}.

\section{Generalization for Ping-Pong Pipelining}

In order to achieve full usage for the off-chip memory bandwidth, we propose to generalize the ping-pong pipelining strategy for arbitan rary number of cores.
Firstly, we would like to quantitatively analyze the utilization for the in~situ write/compute strategy and the naive ping-pong strategy.

\begin{figure}[t]
\includegraphics[width=1\columnwidth]{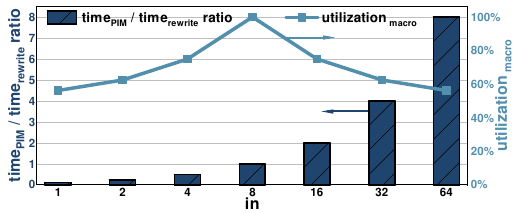}
\vspace{-1.5em}
\caption{the specific idle time ratio of macro}
\label{fig:naive}
\vspace{-1.5em}
\end{figure}

Firstly, we would like to formulate the latency for the memory mode and compute mode.
Given that both weight rewriting and computation are essential operations, we posit that a macro is considered ``idle'' when it is neither performing rewriting nor computation.
When the weight reloading time is less than the PIM time, the rewritten bank has to wait for the PIM bank to finish the computation task of the current layer before starting the computation of the next GeMM operation. 
Assume $size_{macro}$, $size_{OU}$, $n_{in}$, and $s$ represent macro size, operation unit size, number of input vector words for VMM calculaton, and rewrite speed, respectively. During a complete cycle of write and compute, the compute time is: $time_{PIM}=\frac{size_{macro}*n_{in}}{size_{OU}}$
The writing time is $time_{rewrite}=\frac{size_{macro}}{s}$.
When the PIM time is greater than the writing time, the macro utilization is:
\begin{equation}
util_{macro}=\frac{time_{PIM}+time_{rewrite}}{2*time_{PIM}}
\end{equation}
When the PIM time is less than the rewriting time, the macro utilization is: 
\begin{equation}
    util_{macro}=\frac{time_{PIM}+time_{rewrite}}{2*time_{rewrite}}
\end{equation}
With this formulation, Fig.~\ref{fig:naive} shows $time_{PIM}/time_{rewrite}$ ratio and macro utilization for the naive ping-pong strategy under various $n_{in}$ within a specific PIM architecture configuration. In this example, the macro size $size_{macro}$ is set to 32$\times$32 bytes, the output unit size $size_{OU}$ is set to 4$\times$8 bytes, and the bandwidth $s$ is set to 4byte/cycle. It can be observed that only when the number of inputs $n_{in}$ equals 8, where $time_{PIM} = time_{rewrite}$ (i.e. matching the computing time and weight reloading time), at which point the naive ping-pong strategy achieves the highest macro utilization rate. Apart from this scenario, the naive ping-pong strategy significantly reduces macro utilization.

With the aforementioned analysis, in order to maintain the highest macro utilization and off-chip bandwidth utilization during execution for varying values of $n_{in}$, we propose the generalized ping-pong strategy, which directly focuses on the ratio of $time_{PIM}/time_{rewrite}$, and adjusts the start time of each macro execution. This approach averages the demand for off-chip bandwidth across each cycle, thereby reducing the peak demand for off-chip bandwidth. Simultaneously, each macro will immediately transition to the next write/compute operation upon completing the current one, thereby sustaining the highest macro utilization rate.

Fig.~\ref{fig:schedulescheme}(c) illustrates the timing diagram and off-chip memory bandwidth utilization of proposed generalized ping-pong pipeline. \textbf{The core idea of the generalized ping-pong is maintain a peak usage for the off-chip memory bandwidth with multi-core PIM accelerators.} It groups multiple macros for writing and for computing. A deep pipelined pattern is exploited with balanced writing (memory bandwidth occupation) and PIM computing. This scheme has both advantages of in situ write/compute (consistently maintain a high macro utilization rate) and naive ping-pong (keeping high utilization rate for off-chip memory bandwidth).

Assuming the presence of 4 macros in a PIM accelerator, when the ratio of weight updating to computation time is 1:3, \textit{macro2} initiates its weight updating process subsequent to the completion of \textit{macro1}'s rewrite. This sequence continues with \textit{macro3} and \textit{macro4}, effectively distributing the bandwidth demand across each cycle. In this example, compared to the in situ write/compute strategy and the naive ping-pong, the proportion of bandwidth idle time in generalized ping-pong decreased from 75\% and 66\% to 0\%, while the peak bandwidth demand is reduced to 25\% of that required by the in situ write/compute approach. The macro utilization rate in generalized ping-pong remains at 100\%, as the strategy does not induce idle states in the macros. Less bandwidth idle time and higher macro utilization ensure that generalized ping-pong delivers optimal performance under the same bandwidth constraints.

\section{Implement/Deploy Generalized Ping-Pong}
Generalized ping-pong strategy can improve the performance in two cases: (a) \emph{design phase}: design space exploration for full usage of off-chip memory bandwidth in designing a PIM accelerator before tape-out; (b) \emph{runtime phase}: scheduling PIM macros write/compute operations toward the maximum off-chip memory bandwidth utilization after PIM accelerator ASIC fabrication.

\subsection{Design Phase Optimization}
\subsubsection{Synthesizable Base PIM Accelerator Architecture}
To implement the proposed generalized ping-pong, we choose PUMA~\cite{ankit2019puma} design as a synthesizable base PIM accelerator architecture. It executes GeMM computing with compilation optimization. In addition to the original PUMA accelerator design, here we revise the PIM-oriented instruction set architecture (ISA)~\cite{rokicki2018hybrid}. This base architecture supports the aforementioned in~situ write/compute strategy, naive ping-pong strategy and the proposed generalized ping-pong stratety. The ISA comes with an assembler to convert assembly code into binary machine code. The focused scheduling strategies leads to different assembly code for different pipelined execution.

\begin{figure}[t]
\includegraphics[width=1\columnwidth]{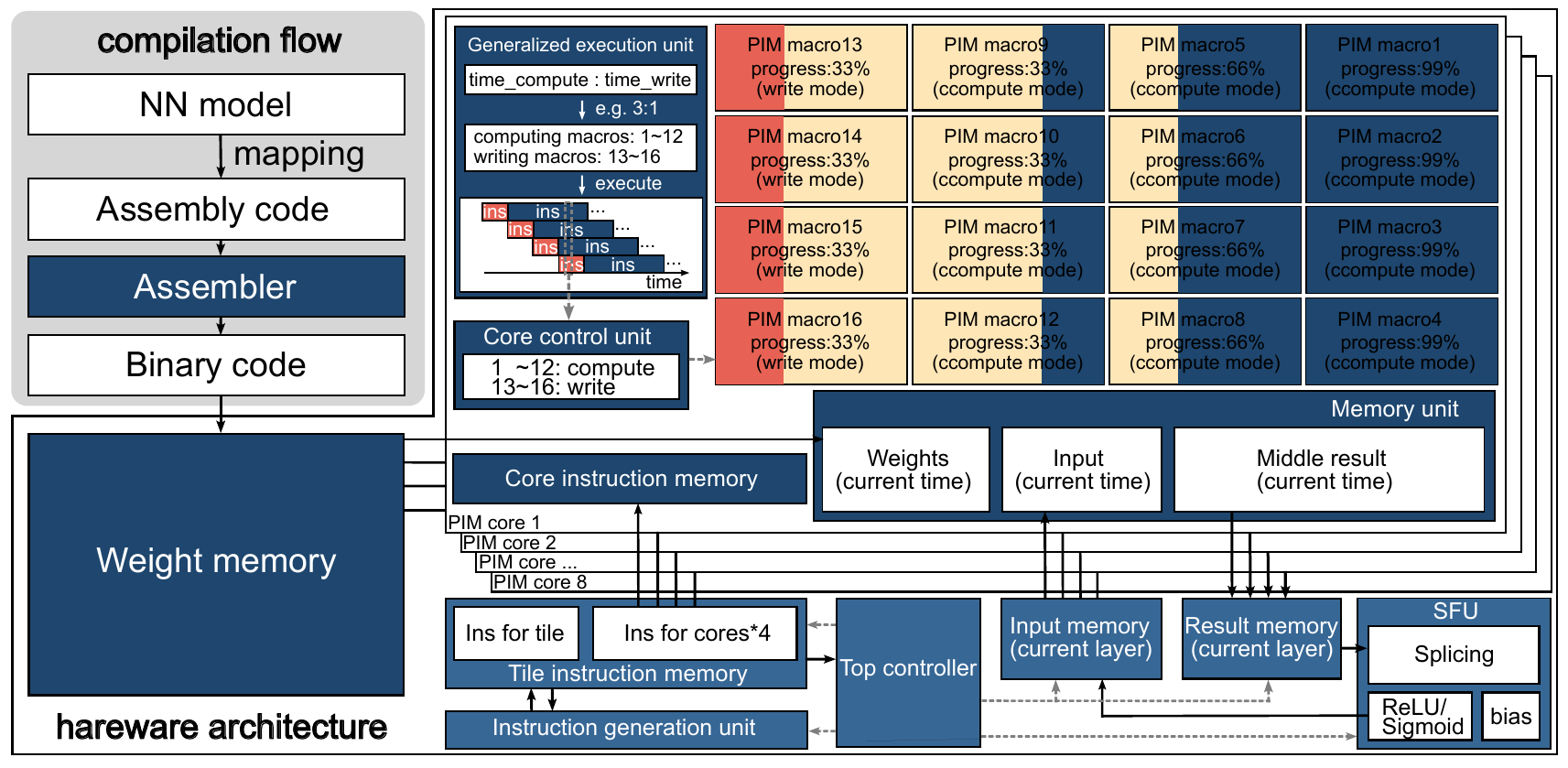}
\vspace{-1.5em}
\caption{The base PIM accelerator architecture as an example to implement generalized ping-pong scheduling strategy. This base architecture is revised from PUMA~\cite{ankit2019puma} to support various scheduling strategies.}
\label{fig:arch}
\vspace{-1em}
\end{figure}

Fig.~\ref{fig:arch} shows the overall revised base architecture. It consists of a global weight memory, a global input memory, a global intermediate result memory, and an instruction memory, which transmit instructions and data between the core and them. The intermediate results are accumulated using vector processing unit (VPU). The architecture also includes a top controller and an instruction generation module. Each PIM core consists of 4 PIM macros, a buffer for storing weights/inputs/intermediate results, a control unit, and core instruction memory. The generalized execution unit is for managing the progress of instruction execution by the core control unit. By allowing or prohibiting the core control unit to operate on specific macros based on the current execution strategy, it enables a specific number of macros to synchronize and perform write/compute operations concurrently.

\subsection{Generalized Ping-Pong in Exploring Design Space}
During the design phase of PIM accelerators, we start from a given off-chip bandwidth and perform the design space exploration with the target generalized ping-pong scheduling strategy. Generalized ping-pong can offer enhanced computational throughput or reduced area overhead. Because of the interruption of PIM computation by weight rewrite, the pursuit is to minimize the number of rewrite operation. Ideally, weights should be written to the macro only once for further reuse. All input vectors should complete VMM with the weights already loaded into the PIM array before the next weight rewrite. However, both input vectors and intermediate result vectors require buffering in on-chip memory. Due to the limited capacity of on-chip memory, the number of input vectors that can be processed at one time is restricted, necessitating the computation of a large number of inputs in batches. This results in a fixed ratio of weight rewrite time to PIM computation time, enabling the integration of the generalized ping-pong strategy into the hardware design phase.

To find the sweet point of 100\% utilization of off-chip memory bandwidth, the design exploration should take generalized ping-pong scheduling into account to match the PIM memory capacity and computing throughput. Table~\ref{tab:1} presents the parameters used in the model. The time of PIM computation is contingent upon the velocity at which the macro completes vector-matrix operations and the number of vectors that need to be computed within a batch. 

In generalized ping-pong, the time for a single weight rewrite is: $ time_{rewrite}=size_{macro}/s$, and the time for a PIM compute is $time_{PIM}=size_{macro}\cdot n_{in}/size_{OU}$.
The number of macros that can be supported under a fixed off-chip bandwidth (with full usage) is given by 
\begin{equation}
num_{macro}=\left\{
\begin{array}{ll}
\dfrac{band.}{s} & \textrm{, in~situ write/compute;}\\
\dfrac{2\times band.}{s} & \textrm{, naive ping-pong.}\\
\end{array}
\right.
\end{equation}
Note in the ping-pong strategy, where macros are divided into two groups that rewrite alternately, the average bandwidth demand per macro is reduced to $(s/2)$.

Generalized ping-pong sets the number of macros that rewrite simultaneously according to the ratio of $time_{rewrite}$ to $time_{PIM}$, with each macro's average bandwidth demand being $\frac{time_{rewrite}*s}{time_{PIM}+time_{rewrite}}$, and the number of macros that can be supported is given by
\begin{equation}
num_{macro}=\frac{(time_{PIM}+time_{rewrite})*band.}{time_{rewrite}*s}
\label{eq:ngpp}
\end{equation}
When the ratio of $time_{rewrite}$ to $time_{PIM}$M is not equal to 1, the naive ping-pong strategy may result in idle states of macros, whereas the in~situ write/compute and generalized ping-pong strategies remain unaffected. As a result, the performance of every macro under the ping-pong strategy reduce to 
$\frac{time_{PIM}+time_{rewrite}}{time_{PIM}+time_{rewrite}+|time_{PIM}-time_{rewrite}|}$
of its original capability.

\begin{table}[t]
\vspace{-0.5em}
\centering
\caption{List of Parameters for Design Space Exploration}
\vspace{-1em}
\begin{tabular}{c|c}
\toprule
 name of parameter & value\\
\hline
$band$ & off-chip bandwidth\\
$size_{macro}$ & macro size\\
$size_{OU}$ & operation unit size\\
$s$ & rewrite speed\\
$n_{in}$ & number of activations for VMM calculaton\\
$time_{PIM}$ & Time of a PIM calculation\\
$time_{rewrite}$ & Time of a weight rewrite\\
$n$ & the multiple of band. reduction\\
$num_{macro}$ & number of macros\\
$m$ & the multiple of $num_{macro}$ reduction\\
\hline
\end{tabular}
\label{tab:1}
\vspace{-1.5em}
\end{table}

Based on the number of macros supported and the performance of each macro, it can be derived that under the current band., the ratio of the number of macros for the three strategies generalized ping-pong:in~situ write/compute: naive ping-pong is 
\begin{equation}
    \frac{size_{macro}*in/size_{OU}+size_{macro}/s}{size_{macro}/s}:1:2
\end{equation}
and the execution time ratio for generalized ping-pong:in~situ write/compute: naive ping-pong is
\begin{equation}
    \frac{in*s+size_{OU}}{size_{OU}}:1:\frac{2*(in*s+size_{OU})}{in*s+size_{OU}+|in*s-size_{OU}|}
\end{equation}

When $time_{PIM}>time_{rewrite}$, the generalized ping-pong strategy demonstrates better performance compared to the other two strategies. When $time_{PIM}<time_{rewrite}$, generalized ping-pong outperforms the in~situ write/compute strategy and offers equivalent performance to the ping-pong strategy while utilizing fewer macros, which translates to a lower area overhead. When $time_{PIM}=time_{rewrite}$, generalized ping-pong provides better performance than the in~situ write/compute strategy, and its performance and number of macros are identical to those of the naive ping-pong strategy. This is because, at this point, the macros in the naive ping-pong strategy do not enter an idle state, and the actual execution methods of the two strategies are completely aligned.

\subsection{Runtime Phase Pipeline Adaption}
In a large system-on-a-chip (SoC) design, the off-chip memory bandwidth for PIM accelerator is often assigned dynamically in runtime. Chances are the accelerator cannot get its full off-chip memory bandwidth. The proposed generalized ping-pong scheduling strategy is also helpful for this case.

For a PIM accelerator after fabrication, when encountering a reduction in off-chip bandwidth during the execution of computational tasks, the generalized ping-pong strategy can preserve a greater portion of performance compared to other strategies. We discuss about the performance degradation caused by the reduction of off-chip bandwidth under the in~situ write/compute, ping-pong, and generalized ping-pong strategies through a modeling approach.

For the in~situ write/compute strategy, when the off-chip bandwidth is reduced to $\frac{band.}{n}$, the optimal response is not to decrease the number of active macros but to reduce the speed of weight updating operations for each macro, thereby lowering the demand for off-chip bandwidth. This means that the number of functioning macros remains constant, but the performance of each macro is diminished. In this case, the performance degradation is: 
\begin{equation}
    \frac{time_{PIM}+time_{rewrite}}{time_{PIM}+time_{rewrite}*n}
\label{eq:perf_de_in_situ}
\end{equation}
In comparison to the strategy of maintaining the speed of weight updating while reducing the number of active macros, which results in performance degradation to $\frac{1}{n}$ of the original case, it can preserve a better proportion of performance.

For the naive ping-pong strategy, when $time_{PIM} > time_{rewrite}$, the response strategy is to maintain the number of active macros and reduce the speed of weight updating operations for each macro to decrease the demand on the off-chip bandwidth. At this point, although $time_{rewrite}$ increases, it still satisfies the condition $time_{PIM} > time_{rewrite}$, which means that the increase in $time_{rewrite}$ merely leads to a reduction in the idle time of the macros, with performance remaining constant until $time_{rewrite}$ increases to the point where $time_{PIM} = time_{rewrite}$. At $time_{PIM} = time_{rewrite}$, each macro's utilization reaches its peak since the macros do not enter an idle state. At this juncture, if the off-chip bandwidth decreases again and $time_{rewrite}$ increases to the point where $time_{PIM} < time_{rewrite}$, the strategy is to maintain the weight updating speed at $time_{PIM} = time_{rewrite}$ and reduce the number of active macros. In this scenario, performance degradation is
\begin{equation}
    \frac{1}{n}.
\label{eq:perf_de_naive}
\end{equation}
Compared to the strategy of reducing the updating speed of weights without decreasing the number of active macros, which results in performance degradation to $\frac{1}{n}$ of the original, the strategy that maintains the speed of weight updating while reducing the number of active macros offers the same performance but with fewer macros in use, thereby reducing energy consumption.

For the generalized ping-pong strategy, when off-chip bandwidth is reduced, the speed of weight updating remains constant while the number of active macros is decreased. Unlike the previous two strategies, generalized ping-pong adjusts the ratio of $time_{PIM}$ to $time_{rewrite}$ to reduce the number of working macros. As previously mentioned, $time_{PIM}$ depends on the speed at which a macro completes vector-matrix operations and the number of vectors that need to be computed within a batch, which is determined by the amount of on-chip memory each macro can access. When the number of working macros is reduced and the on-chip memory capacity remains unchanged, the amount of on-chip memory available to each macro increases, in increases. This implies that $time_{PIM}$ increases. According to Eq.~\ref{eq:ngpp}, when $time_{PIM}$ increases and $time_{rewrite}$ remains constant, it supports a greater number of active macros.

When the off-chip bandwidth is reduced to $band./n$, the number of active macros becomes $num_{macro}/m$ accordingly.
The ratio of $time_{PIM}$ to $time_{rewrite}$ becomes: $\frac{size_{macro}}{size_{OU}}\cdot n_{in}\cdot m:\frac{size_{macro}}{s}$.
At this point, the average demand for off-chip bandwidth per macro is $\frac{time_{rewrite}\cdot s}{time_{PIM}+time_{rewrite}}$
, and multiply it with $\cdot num_{macro}/m$, which should be equal to $band./n$. 
Then we can solve 
the performance degradation:
\begin{equation}
\frac{2(n_{in}*s+size_{OU})}{size_{OU}+\sqrt{size_{OU}^2+\frac{4num_{macro}*size_{OU}*n_{in}*s^2*n}{band.}}}
\label{eq:gppreduce}    
\end{equation}
In Eq~.\ref{eq:gppreduce}, all parameters except for $n$ and $m$ are numerical values obtained during the hardware design phase using the generalized ping-pong strategy.
The slopes of Eq~.\ref{eq:gppreduce}, Eq~.\ref{eq:perf_de_in_situ}, Eq~.\ref{eq:perf_de_naive} demonstrate that the generalized ping-pong strategy can retain a greater portion of performance compared to the other two strategies.

\section{Evaluation}
\label{sec:eval}
\subsection{Experimental Setup}
The proposed generalized ping-pong strategy focuses on the throughput improvement for multi-macro PIM GeMM accelerators. To evaluate it, we implement different accelerator-level concurrent write/computing pipeline strategies on a revised PUMA~\cite{ankit2019puma} design.
To simplify the analysis and control the variables, we focus on large-scale consecutive GeMM operations with basic linear algebra subprograms (BLAS) level benchmarks~\cite{blas}. Because the target pipeline strategy emphasizes the alignment on clock cycles, the timing simulation is based on synthesizable Verilog HDL design (check our open-source repository \url{https://github.com/rw999creator/gpp-pim} to reproduce the simulation results). The example design parameters are set to: PIM accelerator has 16 cores, where each is equipped with 16 macros. The macro size is 32$\times$32 bytes, with a write speed ranging from 1 to 8byte/cycle, and the size of the operating unit is 4$\times$8byte.

\subsection{Evaluation for Design Phase Optimization}

\begin{figure}[t]
	\includegraphics[width=1\columnwidth]{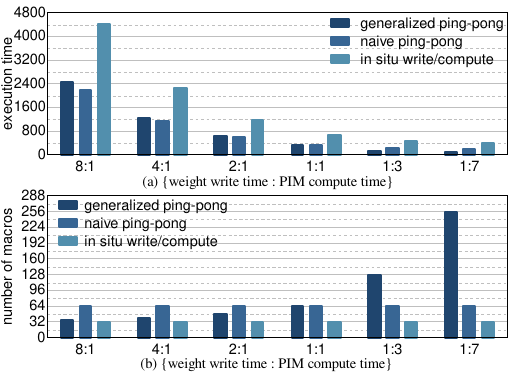}
	\vspace{-1.5em}
	\caption{(a) Execution time comparison under the three strategies. (b) Number of macros comparison under the three strategies. }
	\label{fig:design_space}
	\vspace{-1em}
\end{figure}

Fig.~\ref{fig:design_space} presents a comparison of performance and macro count between the generalized ping-pong and other strategies during the hardware design exploration phase. At this stage, the off-chip bandwidth memory $band.$ is set to 128byte/cycle. The x-axis is the ratio of weight write time\footnote{The ``weight write time'' refers to entirely rewriting the data stored in PIM.} over the PIM compute time. The y-axis is the execution latency in cycle numbers. When $time_{rewrite}<time_{PIM}$, under the same off-chip bandwidth conditions, generalized ping-pong can support a greater computational power compared to the other two strategies, and it requires the use of more macros. In the scenario where the ratio of $time_{rewrite}$ to $time_{PIM}$ is 1:7, generalized ping-pong achieves a 2.51$\times$ performance improvement over naive ping-pong and a 5.03$\times$ improvement over in situ write/compute. When $time_{rewrite}=time_{PIM}$, the generalized ping-pong and naive ping-pong strategies completely overlap, and they exhibit a 2$\times$ performance improvement over in situ write/compute in terms of performance. When $time_{rewrite}>time_{PIM}$, generalized ping-pong outperforms in situ write/compute and matches the performance of naive ping-pong, but with the advantage of using fewer macros, which conserves area and power consumption. In the case where the ratio of $time_{rewrite}$ to $time_{PIM}$ is 8:1, generalized ping-pong reduces the number of macros by 43.75\% compared to naive ping-pong and achieves 1.78$\times$ performance improvement over in situ write/compute strategy. The improvement brought by the generalized ping-pong on performance and area depends on the ratio of $time_{rewrite}$ to $time_{PIM}$.

\vspace{-0.5em}
\subsection{Evaluation for Runtime Phase Adaptation}

\begin{figure}[t]
\includegraphics[width=1\columnwidth]{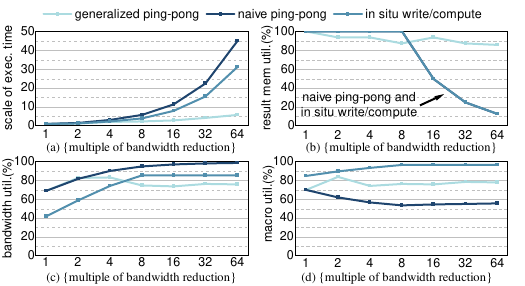}
\vspace{-1.5em}
\caption{(a) Scale of execution time comparison under the three strategies. (b) Result memory utilization comparison under the three strategies. (c) Bandwidth utilization comparison under the three strategies. (d) Macro utilization comparison under the three strategies. }
\label{fig:runtime}
\vspace{-1.5em}
\end{figure}

Fig.~\ref{fig:runtime} shows the results for runtime phase optimization. It shows the comparative performance of the three strategies (in~situ write/compute, naive ping-pong, generalized ping-pong) in response to bandwidth fluctuations. The x-axis is how many times of off-chip memory bandwidth reduction compared to that given during design phase. The y-axes are (a) normalized execution, (b) average on-chip memory utilization rate, (c) off-chip memory bandwidth utilization rate, and (d) average macro utilization rate. For Fig.~\ref{fig:runtime}(a) and (b) This comparison is performed on the design phase optimization goal of $time_{rewrite}$ = $time_{PIM}$ and exerts a progressive reduction in bandwidth to monitor the trend in performance variation. The experimental results indicate that generalized ping-pong can retain a greater degree of performance as off-chip bandwidth decreases, in comparison to current execution schemes. When the bandwidth is reduced to $\frac{band.}{64}$, our strategy achieves 5.38$\times$ improvement in performance over in situ write/compute and 7.71$\times$ improvement in performance over naive ping-pong.

Fig.~\ref{fig:runtime}(c) and (d) show the comparison of off-chip bandwidth utilization and macro utilization rate, respectively. The in situ write/compute strategy yields a lower off-chip bandwidth utilization, whereas the naive ping-pong strategy has a lower macro utilization. The advantage of generalized ping-pong is with both high off-chip bandwidth utilization and macro utilization.

Table~\ref{tab:2} shows the design space optimization with generalized ping-pong at different off-chip bandwidth (unit: byte/cycle). The discrepancy between the execution strategies calculated by the model (with a fractional number of PIM macros) and those actually implemented in Verilog HDL (with integer number of PIM macros) diminishes as the number of macros increases. For the in situ write/compute strategy, the optimal scheduling is to reduce the speed at which each macro rewrites weights, thereby decreasing the demand for off-chip bandwidth, while keeping the number of active macros constant. However, the speed of weight updating cannot be infinitely reduced as a latency overhead. When the speed of weight updating reaches the minimum value determined by hardware design, it becomes necessary to reduce the number of active macros to cope with further decreases in bandwidth. This leads to a more rapid decline in performance. For generalized ping-pong, due to the finite number of macros, the actual execution results are an approximation of the model. 
\begin{table}[h]
\vspace{-1em}
\centering
\caption{The discrepancy between theory and practice}
\vspace{-1em}
\begin{tabular}{c|c|c|c|c|c|c}
\hline
\multirow{2}{*}{$band.$} & \multicolumn{2}{c|}{working macros} & \multicolumn{2}{c|}{time\_PIM:time\_rew} & \multicolumn{2}{c}{remaining perf.} \\
\cline{2-7}
& theory & practice & theory & practice & theory & practice\\
\hline
256 & 82.05 & 80 & 1.56:1 & 1.5:1 & 78.08\% & 75.00\%\\
128 & 54.01 & 49 & 2.37:1 & 2.5:1 & 59.31\% & 54.69\%\\
64  & 36.26 & 36 & 3.53:1 & 3.5:1 & 44.14\% & 43.75\%\\
32  & 24.71 & 24 & 5.18:1 & 5:1 & 32.37\% & 31.25\%\\
16  & 17.02 & 16 & 7:52:1 & 7:1 & 23.49\% & 21.88\%\\
8   & 11.83 & 11 & 10.82:1 & 10:1 & 16.91\% & 15.63\%\\
\hline
\end{tabular}
\label{tab:2}
\vspace{-1em}
\end{table}

\section{Conclusion}
This paper attempts to answer the question of \emph{how to realize concurrent weight trasnfer and PIM computation towards upscaled GeMM operations}.
To achieve this, we propose a novel generalized ping-pong pipelining strategy for arbitary scale of PIM accelerators. With an exemplary PIM accelerator implemented, we demonstrate the efficacy of the generalized ping-pong strategy. It is applicable for design space exploration and improve runtime off-chip memory bandwidth utilization. Compared to existing strategies, our approach achieves superior performance boost under the same off-chip memory bandwidth. This work reveils the fundamental theory of pipeling optimization for PIM architectures.


\end{document}